\documentclass[prb,twocolumn,showpacs,floatfix,amsmath,amssymb,superscriptaddress]{revtex4-1}
\usepackage{amsfonts}
\usepackage{stmaryrd}
\usepackage{bbm}
\usepackage{mathrsfs}
\usepackage{tipa}
\usepackage{amssymb}
\usepackage{txfonts}
\usepackage{graphicx}
\usepackage{dcolumn}
\usepackage{epstopdf}
\usepackage[colorlinks,linkcolor=blue,urlcolor=blue,citecolor=blue]{hyperref}
\usepackage{multirow}
\usepackage{subfigure}
\usepackage{url}
\usepackage{upgreek}
\usepackage{siunitx}
\usepackage[utf8]{inputenc}
\usepackage[english]{babel}
\usepackage{xcolor}

\begin{document}

\title{Relaxation dynamics of the optically driven nonequilibrium states in the electron- and hole-doped topological-insulator materials (Bi$_{1-x}$Sb$_x$)$_2$Te$_3$}

\author{Chris Reinhoffer}
\affiliation{Institute of Physics II, University of Cologne, 50937 Cologne, Germany}

\author{Yu Mukai}
\affiliation{Institute of Physics II, University of Cologne, 50937 Cologne, Germany}
\affiliation{Department of Electronic Science and Engineering, Kyoto University, Kyoto, Japan}

\author{Semyon Germanskiy}
\affiliation{Institute of Physics II, University of Cologne, 50937 Cologne, Germany}

\author{Andrea Bliesener}
\affiliation{Institute of Physics II, University of Cologne, 50937 Cologne, Germany}

\author{Gertjan Lippertz}
\affiliation{Institute of Physics II, University of Cologne, 50937 Cologne, Germany}
\affiliation{KU Leuven Quantum Solid State Physics, Celestijnenlaan 200 D, 3001 Leuven, Belgium}

\author{Anjana Uday}
\affiliation{Institute of Physics II, University of Cologne, 50937 Cologne, Germany}

\author{A. A. Taskin}
\affiliation{Institute of Physics II, University of Cologne, 50937 Cologne, Germany}

\author{Yoichi Ando}
\affiliation{Institute of Physics II, University of Cologne, 50937 Cologne, Germany}

\author{Zhe Wang}
\affiliation{Institute of Physics II, University of Cologne, 50937 Cologne, Germany}
\affiliation{Fakultät Physik, Technische  Universität  Dortmund, 44227 Dortmund, Germany}

\author{Paul H. M. van Loosdrecht}
\affiliation{Institute of Physics II, University of Cologne, 50937 Cologne, Germany}

\date{\today}

\begin{abstract}
We report on time-resolved mid-infrared-pump terahertz-transmission-probe studies of the topological-insulator materials (Bi$_{1-x}$Sb$_{x}$)$_2$Te$_3$, in which by varying $x$ charge carriers are chemically tuned to be of \textit{n}-type or \textit{p}-type.
Relaxation dynamics is found to be different in various aspects for transitions below or above the bandgap, which are selectively excited by changing the pump-pulse energy.
For the below-bandgap excitation, an exponential decay of the pump-probe signals is observed, which exhibits linear dependence on the pump-pulse fluence.
In contrast, the relaxation dynamics for the above-bandgap excitation is characterized by a compressed exponential decay and nonlinear fluence dependence at high pump flunences, which reflects interaction of the excited nonequilibrium states.
\end{abstract}

\maketitle

\section{Introduction}

Topological quantum materials, such as topological insulators (TIs) and Dirac or Weyl semimetals, have recently been extensively investigated due to their highly nontrivial properties, e.g. topological surface states, linear dispersion relation, and protection from back scattering, which make them potentially applicable for new technologies. \cite{Qi2010a,Hasan2010,Moore2010,Ando2013,RevModPhys.90.015001}.
At the same time, unconventional physical phenomena have been observed in light driven nonequilibrium states \cite{Wang2013,Giorgianni2016,McIver2019,RMAD2020}, in which the dynamical properties have been relatively less investigated \cite{Oka19}.\\
\indent A direct probe of the time-dependent evolution of the surface states in a nonequilibrium TI system can be provided by time- and angle-resolved electron emission spectroscopy (trARPES) \cite{Sobota2012,Hajlaoui2012,Crepaldi2012,Wang2012,Hajlaoui2014,Neupane2015}. Apart from the bulk states as in a conventional insulator, the surface states act as an additional channel for the system to relax back to the equilibrium state. 
Depending on the specific band structures, the lifetime of the nonequilibrium states in a TI can vary from a few picoseconds (ps), e.g. in (Bi$_{0.2}$Sb$_{0.8}$)$_2$Te$_3$, to the order of microseconds ($\mu$s) such as in Bi$_2$Te$_2$Se \cite{Neupane2015}.
Nonequilibrium dynamics of charge carriers in TIs has also been studied using time-resolved optical spectroscopic techniques.
Optical-pump THz-probe (OPTP) measurements in Bi$_{2}$Se$_{3}$ thin films
revealed a negative change of low-frequency THz optical conductance which was attributed to the metallic response of the surface states \cite{Sim2014}.
By studying thickness dependence of the OPTP response in Bi$_{2}$Se$_{3}$ thin films, 
a shorter time scale of $\sim 5$~ps was observed for thicker films which was ascribed to the dynamics of bulk charge carriers, whereas a longer time scale of $\sim 10$~ps in thinner films was assigned to surface states \cite{Aguilar2015}. 
Using mid-infrared-pump THz-probe and THz-pump THz-probe spectroscopy on Bi$_{2}$Se$_{3}$ thin films, a negative change of the THz optical conductivity was observed for both situations \cite{Luo2019}.
These results were explained by the co-existence of two types of electron plasmas, one for bulk and the other for surface states \cite{Luo2019}. 
A non-linear dependence of the pump-induced change of THz electric field transmission of pump power was observed in the bulk-insulating TI Bi$_{1.5}$Sb$_{0.5}$Te$_{1.7}$Se$_{1.3}$, which was related to the relaxation through the interaction between bulk and surface states, in particular, the injection of photo induced bulk carriers into the surface states\cite{Choi2018}.\\
\indent Here, using time-resolved mid-infrared-pump THz-transmission-probe spectroscopy, we study the nonequilibrium electronic dynamics in the TI systems (Bi$_{1-x}$Sb$_{x}$)$_2$Te$_3$ (BST) as a function of temperature and pump-pulse fluence. By varying the content of Sb, the charge carriers of the system bulk can be tuned to be of electron-type (\textit{n}-type) or of hole-type (\textit{p}-type) \cite{Zhang2011}.
The nonequilibrium dynamics is further investigated by changing the pump-pulse energies which allows a selective excitation below or above the bandgap. 
While for the below-bandgap excitation the relaxation dynamics exhibits an exponential decay as usually expected for a noninteracting system, a clear deviation from the exponential decay is revealed for the above-bandgap excitation.
Phenomenologically, the observed relaxation behavior can be described by a compressed exponential decay, hinting at strong interaction of nonequilibrium states\cite{Izrailev2006}.
In addition, our measurements reveal a non-linear dependence on pump-pulse fluence for the above-bandgap excitation, in contrast to the linear dependence observed in the below-bandgap excitation measurements.

\begin{figure}[t]
        \center{\includegraphics[width=0.45\textwidth]
        {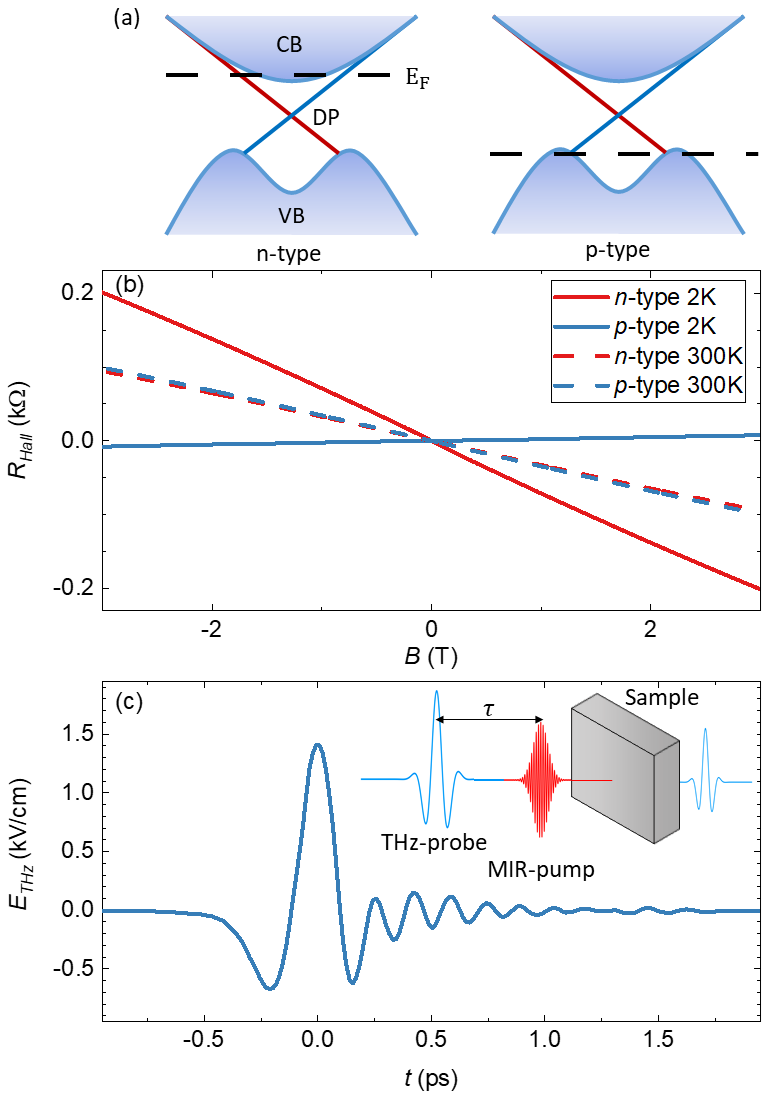}}
        \caption{\label{fig1} (a) Schematic picture of band structure for the \textit{n}- and \textit{p}-type samples with CB, DP, VB and E$_F$, being the conduction band, Dirac point, valence band and Fermi energy, respectively. (b) Hall measurements of the \textit{n}- and \textit{p}-type BST samples at \SI{2}{K} and \SI{300}{K}. (c) Electric field of the THz probe pulse in the time domain, the inset illustrates the MIR pump THz probe experiment.}
\end{figure}

\section{Experimental details}
The BST thin films for our spectroscopic studies were grown on sapphire substrates by molecular beam epitaxy (MBE) and characterized by transport measurements \cite{Yang2014}, to extract information about the charge carrier type and density.
The majority charge carriers are electrons or holes which are indicated by the negative or positive slope in the Hall measurements\cite{Yang2014}, respectively, for the \textit{n}- or \textit{p}-type samples at \SI{2}{K}, as shown in Fig. \ref{fig1} (b).
These features are illustrated in the band structures in Fig.~\ref{fig1} (a).
The thickness of the \textit{n}-type and the \textit{p}-type samples was \SI{14}{nm} and \SI{15}{nm}, respectively.
The electronic bangap in both samples was $\approx$ \SI{200}{meV} at \SI{300}{K} \cite{Ando2013,Yang2014, Zhang2011}.
To investigate the nonequilibrium dynamics, we performed mid-infrared pump THz-probe measurements in a transmission geometry as depicted in the inset of Fig. \ref{fig1} (c).
The THz-pulses were generated by optical rectification\cite{Ulbricht2011} of a \SI{800}{nm} pulses at a repetition rate of \SI{1}{kHz} from an amplified laser system in a \SI{0.3}{mm} thick GaP crystal, and detected by electro-optic sampling \cite{Ulbricht2011} in a GaP crystal with the same thickness. 
Time dependence of the electric field of the THz probe-pulse is shown in Fig.~\ref{fig1} (c) with a maximum electric field strength of $\approx$ \SI{1.5}{kV/cm}. 
The complete THz beam path was kept inside vacuum to avoid water absorption.
In our experiment the pump pulses with a central energy of \SI{150}{meV} and \SI{500}{meV} were used for excitations below and above the bandgap, respectively. The \SI{500}{meV} was generated from an optical parametric amplifier (OPA), while the \SI{150}{meV} was generated by noncollinear difference frequency mixing using the signal and idler output of the OPA.
Maximum powers of \SI{0.4}{mW} and \SI{16}{mW} with the temporal width of
$\approx$ \SI{850}{fs} and $\approx$ \SI{120}{fs} were achieved for the \SI{150}{meV} and \SI{500}{meV} pump pulses, respectively.
The pump beam was modulated by a chopper for lock-in amplification, and the change of the peak electric field of the THz pulse was measured as a function of pump-probe time delay.
The samples were kept in a continuous flow cryostat for a temperature range from \SI{4.2}{} to \SI{300}{K}.

\section{Experimental results}

\subsection{\textit{n}-type (Bi$_{1-x}$Sb$_x$)$_2$Te$_3$}

\begin{figure}[t]
        \center{\includegraphics[width=0.45\textwidth]
        {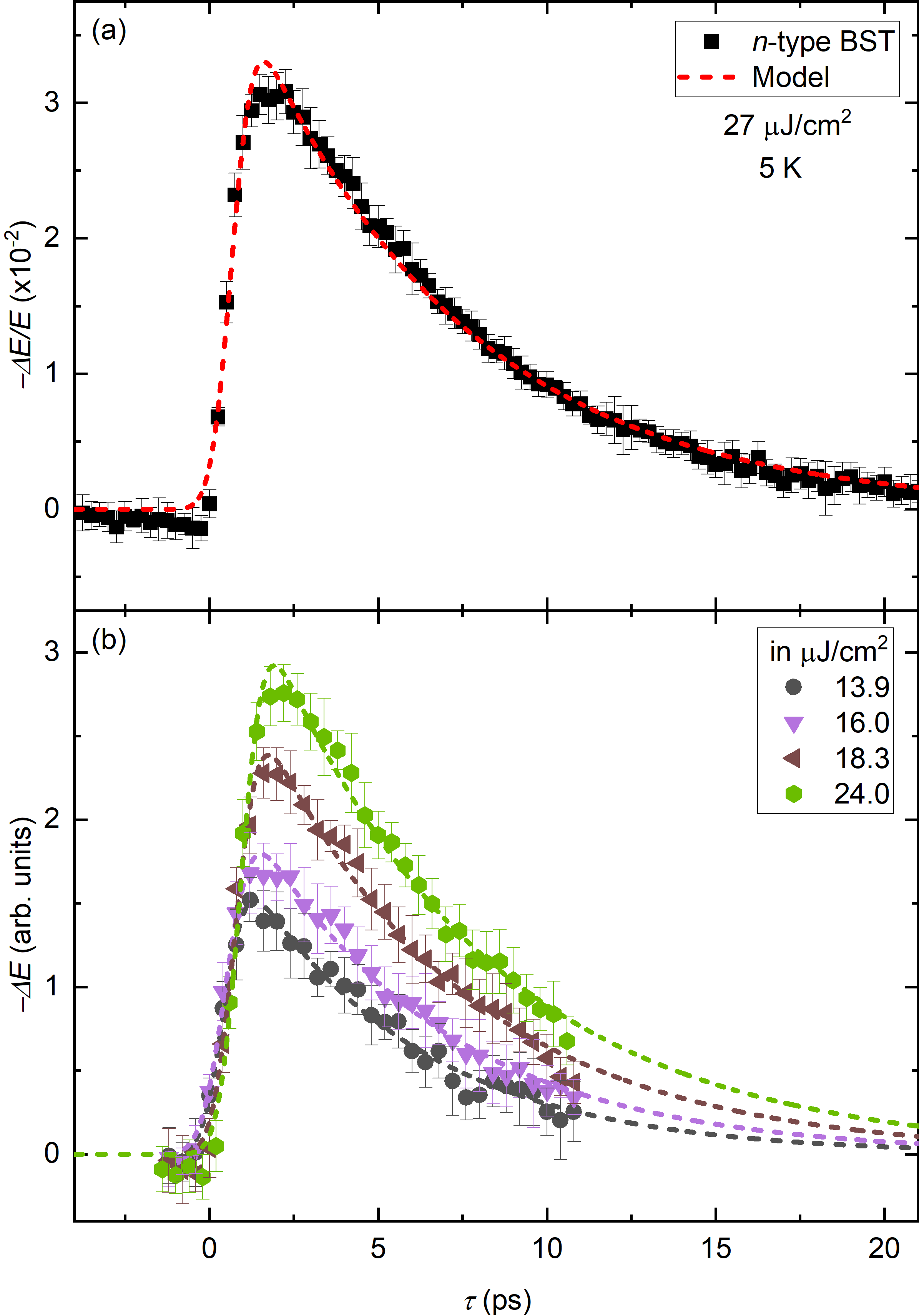}}
        \caption{\label{fig2} Pump-probe measurements on the \textit{n}-type sample with \textit{below-bandgap} excitation at \SI{5}{K}. (a) Complete pump-probe trace at \SI{27}{\micro J/cm^2} (symbols). The dashed line is a fit to a two-level system as described by Eq. \ref{func1} and \ref{func2}. (b) Fluence dependence at \SI{5}{K} for four different fluences, the two-level-system fits (dashed lines) are shown as dashed lines with the same color as the corresponding data (symbols).}
\end{figure}

\begin{figure}[t]
        \center{\includegraphics[width=0.45\textwidth]
        {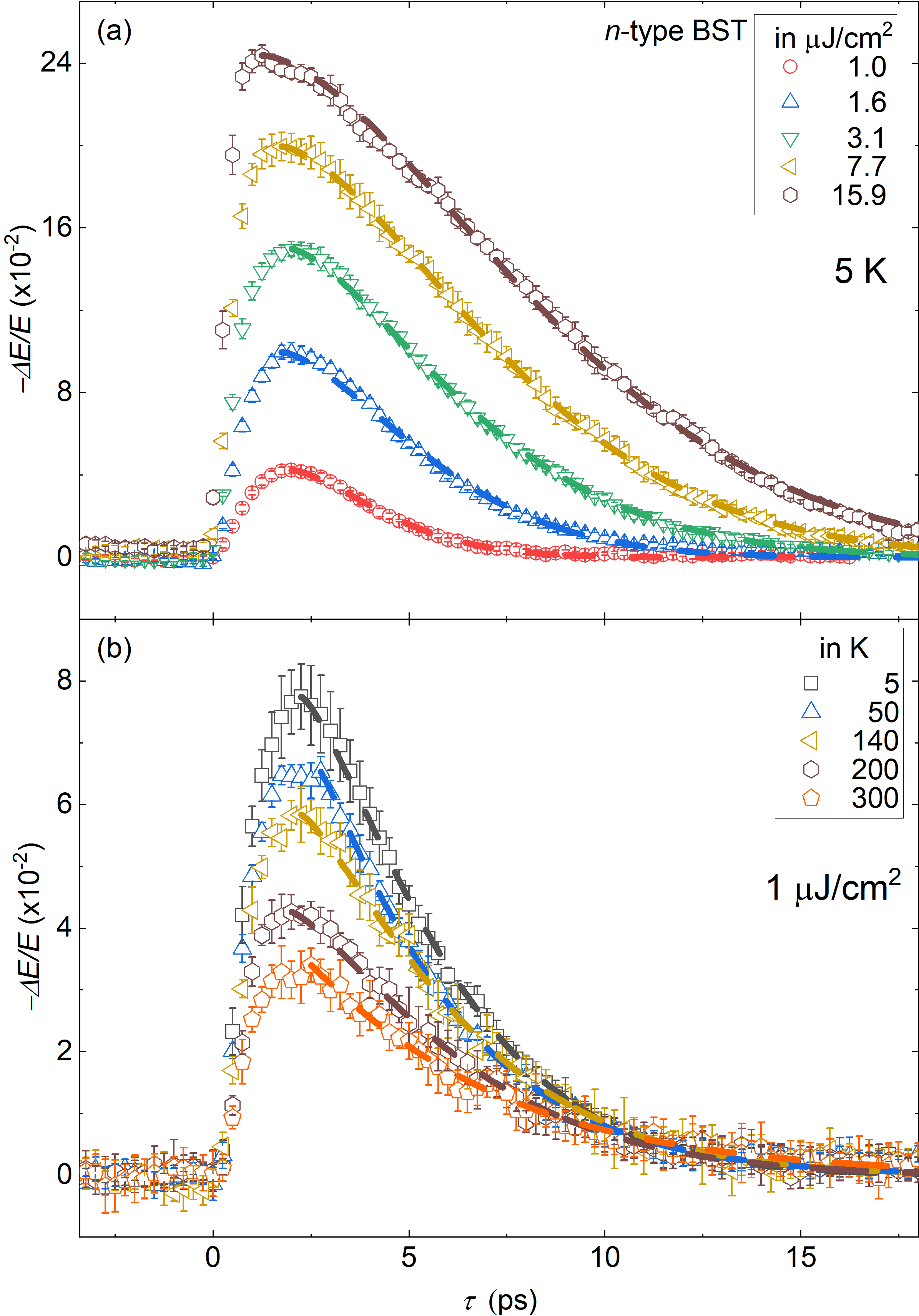}}
        \caption{\label{fig3} \textit{Above-bandgap} excitation measurements on the \textit{n}-type sample. The dashed lines are fits with the compressed exponential function. (a) Fluence dependence measurements at \SI{5}{K} with the corresponding fits as dashed lines. (b) Temperature dependence measurement at \SI{1}{\micro J/cm^2} with the corresponding fits as dashed lines.}
\end{figure}

Figure~\ref{fig2} presents the results for the below-bandgap excitation in the \textit{n}-type sample. Figure~\ref{fig2} (a) shows $-\Delta E/E$, the pump-induced change of the peak THz electric field $\Delta E$ relative to the unperturbed THz field $E$, as a function of the time delay $\tau$ between the mid-IR pump pulse and THz probe pulse, measured at \SI{5}{K} with a pump fluence of \SI{27}{\micro J/cm^2}.
Upon pump-probe overlap at $\tau = 0$ the peak THz electric field is reduced up to a change of \SI{-3}{\percent} at $\tau = \SI{2}{ps}$.
After this minimum, the signal relaxes towards its equilibrium value in about \SI{20}{ps}, following an exponential decay.
Figure~\ref{fig2} (b) shows fluence dependence measurements at \SI{5}{K}.
With decreasing pump fluence from \SI{24}{} to \SI{13.9}{\micro J/cm^2}, the maximum change of $-\Delta E$ decreases by a factor of 2.
This dependence often is observed in pump-probe experiments because the change of the THz electric field typically depends on the number of excited charged carriers\cite{Ulbricht2011}.\\
\indent Figure \ref{fig3} shows the pump-probe measurements on the \textit{n}-type sample for the \textit{above-bandgap} excitation.
The maximum of $-\Delta E/E$ reaches up to \SI{24}{\percent} at a fluence of \SI{15.9}{\micro J/cm^2}.
Even at a much lower fluence of \SI{1}{\micro J/cm^2} there is still a change of \SI{4}{\percent}, exceeding the \SI{3}{\percent} change of the below-bandgap measurement at the highest fluence of \SI{27}{\micro J/cm^2} (see Fig.~\ref{fig2}).
In contrast to the results with below-bandgap excitation, the signal exhibits a broader maximum and at high fluences (Fig.~\ref{fig3} (a)) a change from an exponential decay to a slower decay. 
Furthermore, the maximum of $-\Delta E/E$ and the relaxation time both increase with increasing fluence, as observed in the below-bandgap measurements (Fig.~\ref{fig2} (b)).
Figure~\ref{fig3} (b) shows the temperature dependence of the pump-probe signal at a fluence of \SI{1}{\micro J/cm^2}.
The maximum of $-\Delta E/E$ decreases from \SI{8}{\percent} to \SI{3}{\percent} with increasing temperature from \SI{5}{K} to \SI{300}{K}.
Noticeably, the relaxation in the above-bandgap measurement, in the lower fluence region, is almost twice faster, than in the below-bandgap measurement.

\subsection{\textit{p}-type (Bi$_{1-x}$Sb$_x$)$_2$Te$_3$}

\begin{figure}[t]
        \center{\includegraphics[width=0.45\textwidth]
        {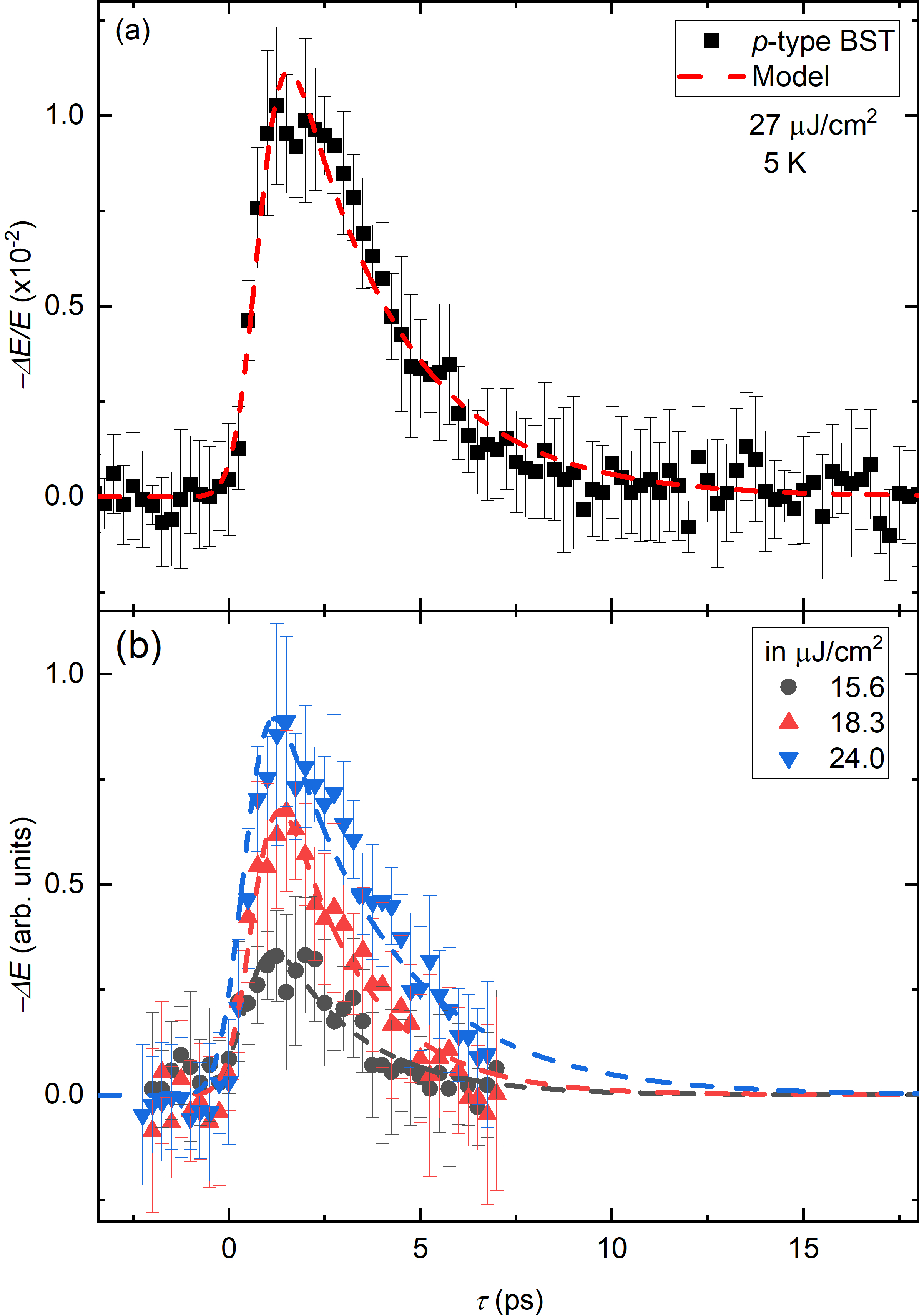}}
        \caption{\label{fig4} Pump-probe measurements on the \textit{p}-type sample with \textit{below-bandgap} excitation at \SI{5}{K}. (a) Complete pump-probe trace at \SI{27}{\micro J/cm^2}, black squares and red dashed line are the measured data and a two-level-system fit, described by Eq.~(\ref{func1}) and (\ref{func2}), respectively. (b) Fluence dependence \SI{5}{K} for three different fluences, the two level fits are shown as dashed lines with the same color as the corresponding data.}
\end{figure}

Figure \ref{fig4} (a) shows the pump-probe measurement on the \textit{p}-type sample for the \textit{below-bandgap} excitation at \SI{5}{K} and \SI{27}{\micro J/cm^2}.
Comparing this to the below-bandgap measurement of the \textit{n}-type sample (Fig. \ref{fig2}) shows differences in the relaxation behavior.
Firstly, the \textit{p}-type system relaxes within about \SI{10}{ps}, while for the \textit{n}-type sample it takes about \SI{20}{ps}.
Secondly, at the highest fluence of \SI{27}{\micro J/cm^2}, the maximum induced change of the electric field is \SI{1}{\percent} in the \textit{p}-type sample, hence evidently lower than the \SI{3}{\percent} change in the \textit{n}-type sample.
Despite these differences, the \textit{p}-type sample (Fig.~\ref{fig4} (b)) shows a similar dependence on the fluence as the \textit{n}-type sample, i.e. $-\Delta E$ and the relaxation time increase with increasing fluence.

\begin{figure}[t]
        \center{\includegraphics[width=0.45\textwidth]
        {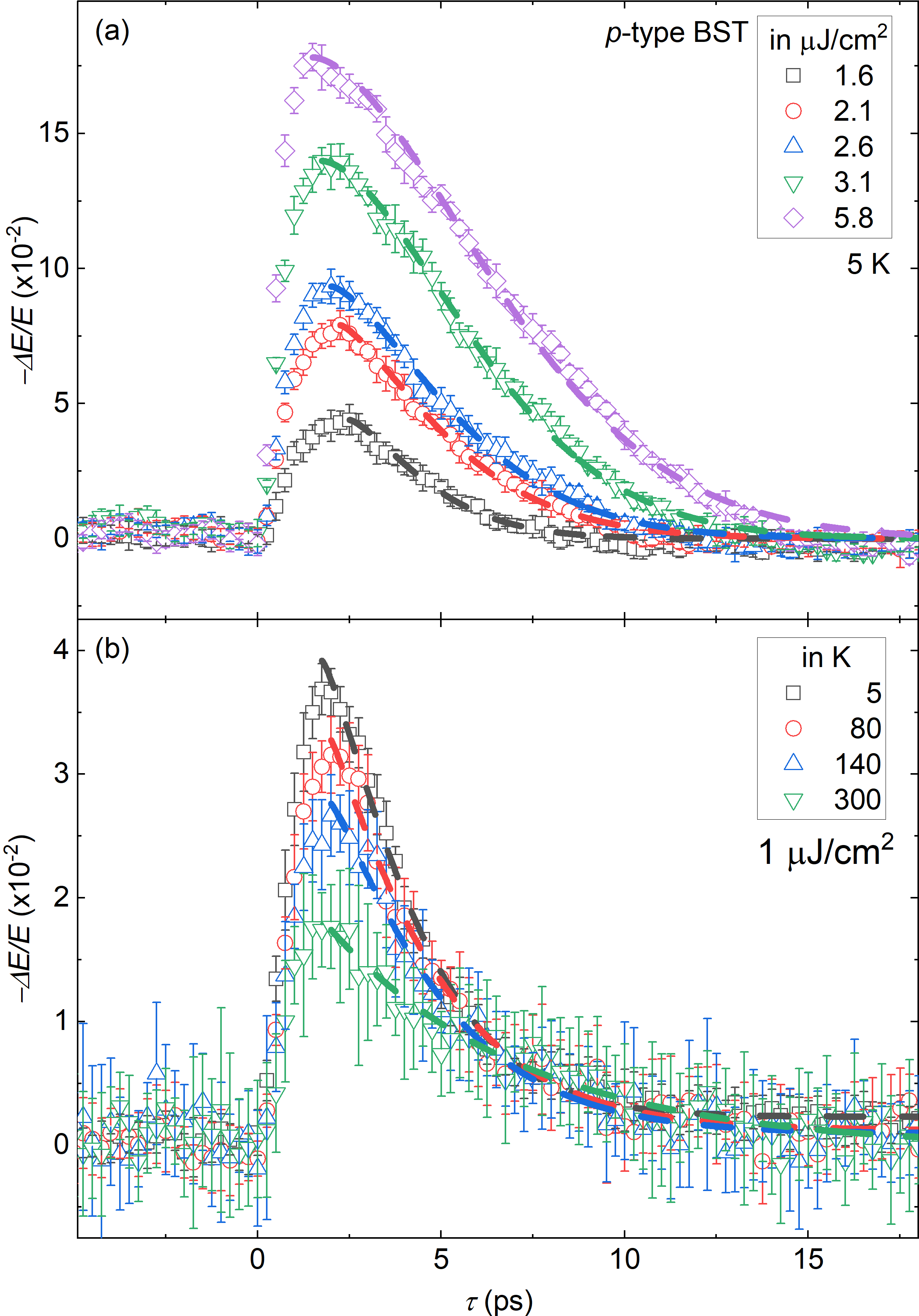}}
        \caption{\label{fig5} \textit{Above-bandgap} excitation measurements on the \textit{p}-type sample. The dashed lines are fits with the compressed exponential function. (a) Fluence dependence measurements at \SI{5}{K} with the corresponding fits as dashed lines. (b) Temperature dependence measurement at \SI{1}{\micro J/cm^2} with the corresponding fits as dashed lines.}
\end{figure}

Figure \ref{fig5} shows the pump-probe measurement on the \textit{p}-type sample for \textit{above-bandgap} excitation.
The fluence dependence measurements in Fig.~\ref{fig5} (a) show a similar change in signal shape from exponential to a slower decay as the measurements on the \textit{n}-type sample (Fig.~\ref{fig3} (a)). 
With increasing fluence from \SI{1.6}{\micro J/cm^2} to \SI{5.8}{\micro J/cm^2},
the signal amplitude increases from \SI{4}{\percent} to \SI{18}{\percent}, while the relaxation time increases from $\approx \SI{10}{ps}$ to $\approx \SI{15}{ps}$.
Figure~\ref{fig5} (b) shows that the pump-probe signal decreases from \SI{4}{\percent} to \SI{1.5}{\percent} with increasing temperature from 5 to 300~K.
Interestingly, the \textit{p}-type sample relaxes within $\approx \SI{10}{ps}$ in the above-bandgap excitation, which is a similar value as in the below-bandgap excitation.
This is in contrast to the \textit{n}-type sample, where the above-bandgap excitation measurement relaxes within $\approx \SI{10}{ps}$ and the below-bandgap measurement within $\approx \SI{20}{ps}$ (see Fig. \ref{fig2} and \ref{fig3}).\\

\section{Discussion}

\begin{figure}[t]
        \center{\includegraphics[width=0.45\textwidth]
        {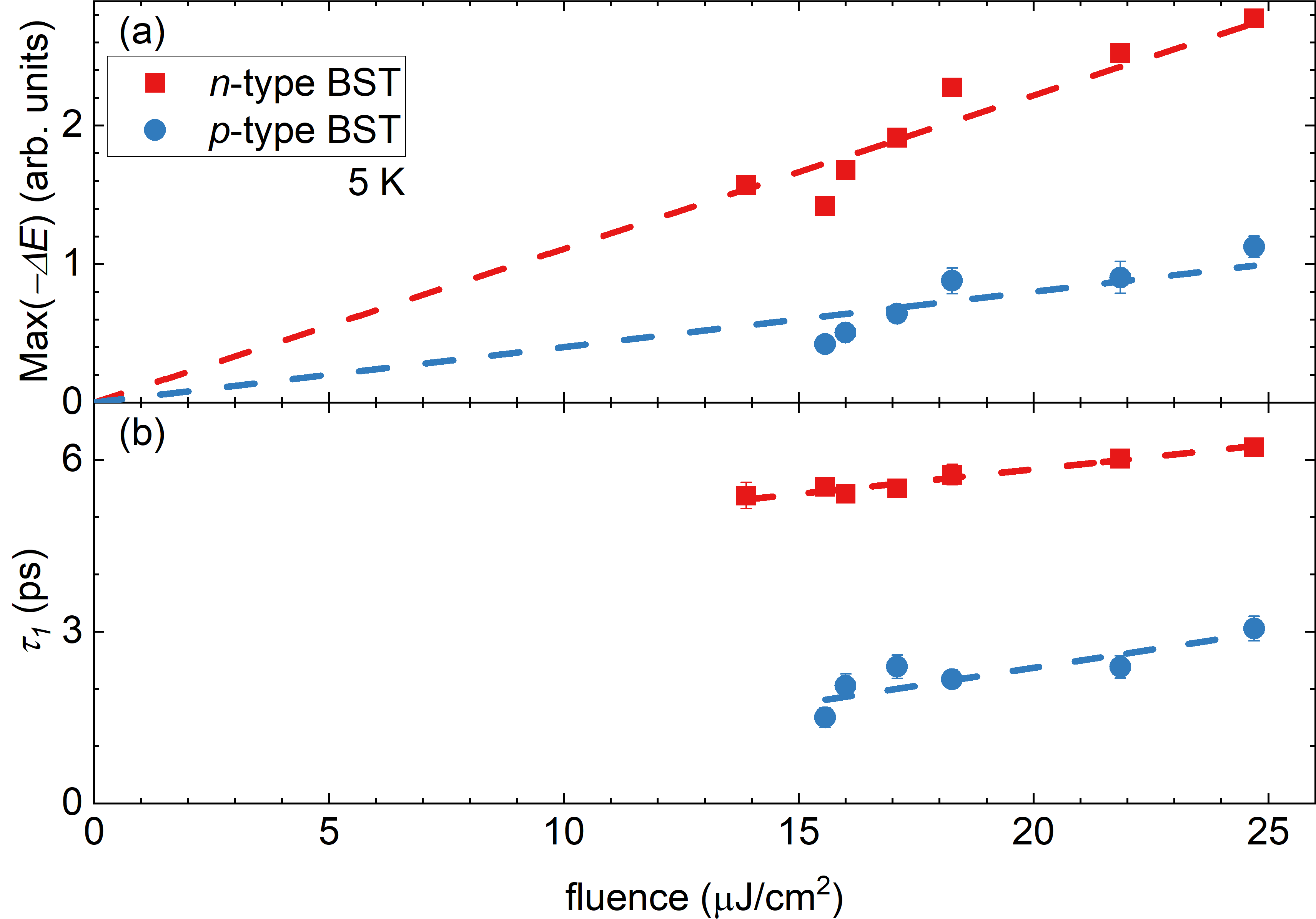}}
        \caption{\label{fig6} Parameters of the fits on the fluence dependence measurements for the \textit{below-bandgap} measurements at \SI{5}{K} (see Figs. \ref{fig2} and \ref{fig4}). (a)  Max($-\Delta E$) with the measured data in red squares and blue dots. The dashed lines are guide for the eye. (b) The relaxation time is linearly dependent on the fluence, the \textit{p}-type sample relaxes twice as fast as the \textit{n}-type sample.}
\end{figure}

\begin{figure}[t]
        \center{\includegraphics[width=0.45\textwidth]
        {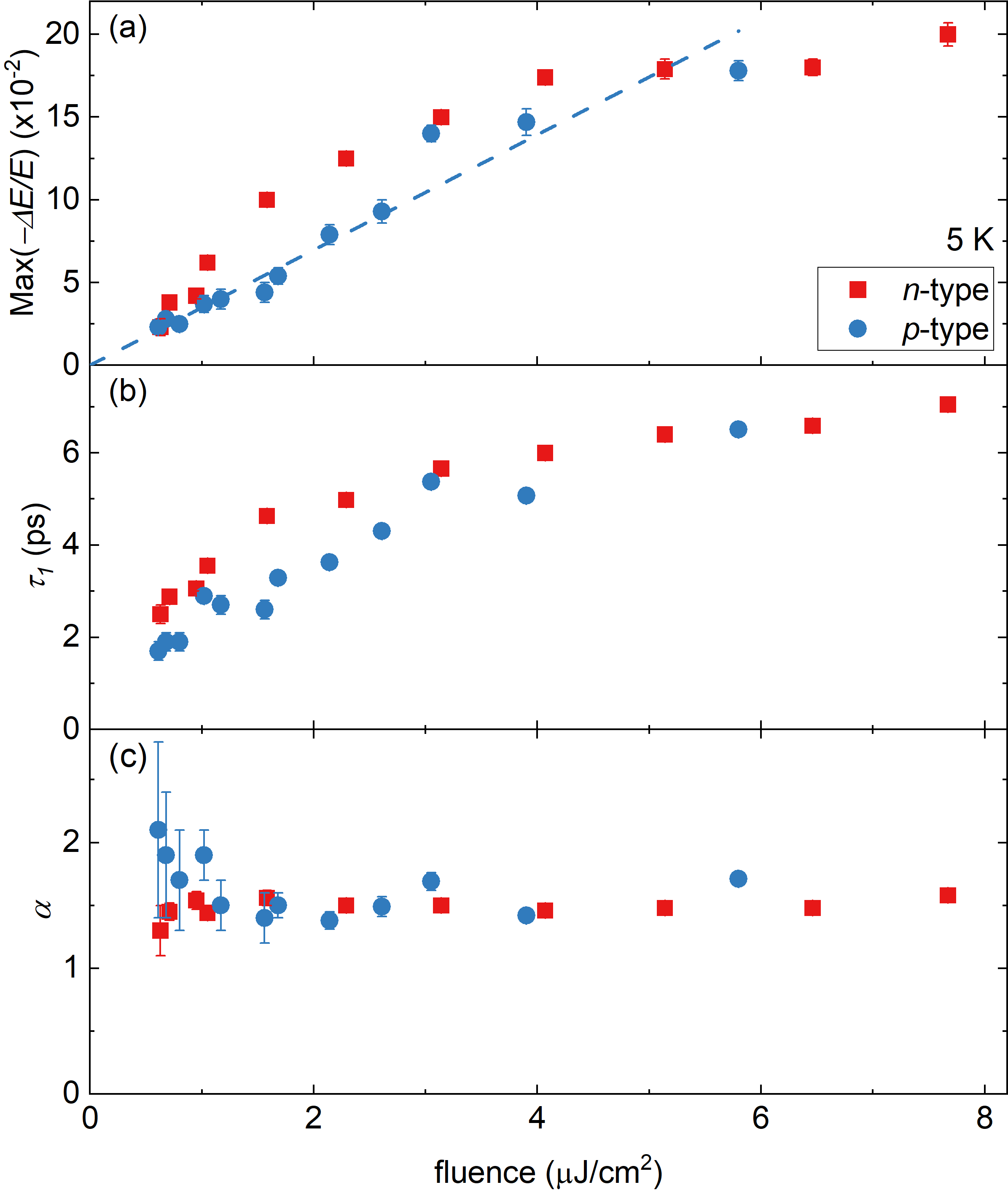}}
        \caption{\label{fig7} Parameter obtained from the compressed exponential fit (Eq. \ref{func3}) of the \textit{above-bandgap} excitation as a function of fluence for the \textit{n}- and \textit{p}-type sample (see Fig. \ref{fig3} and \ref{fig5}). (a) The \textit{p}-type sample shows clear linear dependence on the fluence, while the \textit{n}-type sample shows a nonlinear dependence. (b) The relaxation time shows a nonlinear dependence on the fluence and is similar in magnitude for both samples. (c) The compression parameter is constant within its uncertainty over the whole measurement at a value of $\simeq \SI{1.5}{}$ for both samples.}
\end{figure}

\begin{figure}[t]
        \center{\includegraphics[width=0.45\textwidth]
        {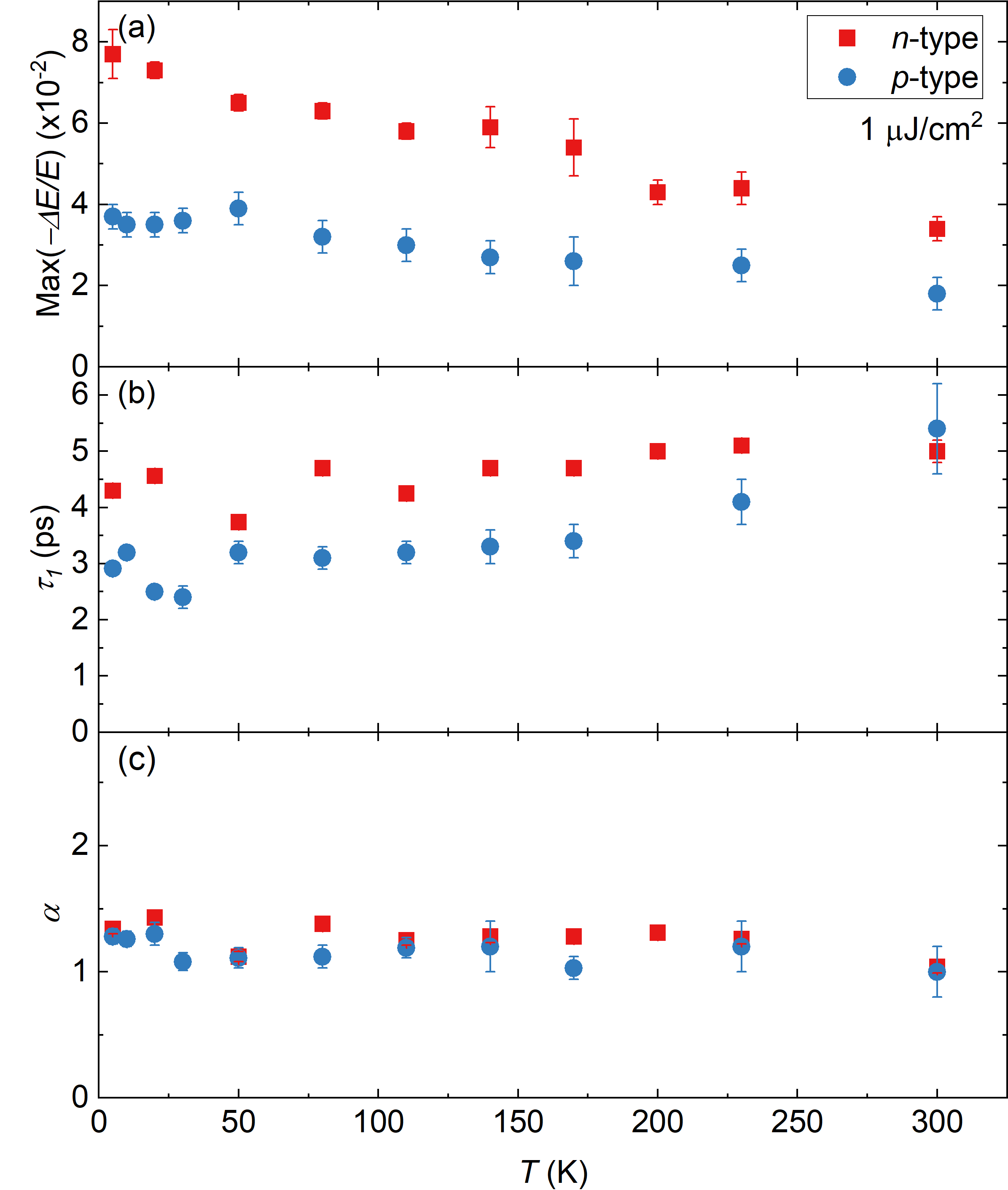}}
        \caption{\label{fig8}  Parameter obtained from the compressed exponential fit (Eq. \ref{func3}) of the \textit{above-bandgap} excitation as a function of temperature for the \textit{n}- and \textit{p}-type sample (see Fig. \ref{fig3} and \ref{fig5}). (a) Max($-\Delta E/E)$ decreases linear with increasing temperature with a different slope for both samples, with the \textit{n}-type sample showing the stronger decrease. (b) The relaxation time shows almost no temperature dependence in both samples. (c) The compression parameter shows a slight decrease with increasing temperature.}
\end{figure}

To quantify the differences for the different types of samples and the different excitation energies (see Fig.~\ref{fig2}-\ref{fig5}), we use phenomenological models to fit the results.
In the below-bandgap excitation measurements, one can clearly observe an exponential decay, while in the above-bandgap measurements the decay behavior deviates from an exponential one.
As shown in Fig.~\ref{fig2} and \ref{fig4}, the below-bandgap excitation measurements are well described by a system which relaxes exponentially. 
Max$(-\Delta E)$ in Fig.~\ref{fig6} (a) is the maximum value directly read from the experimental data as a function of pump fluence.
The fluence dependence exhibits a quasi-linear behavior, 
indicating that the excitation process is a linear one-photon process.
Based on the relaxation, the linearity of the excitation, and the expected underlying band structure of BST we adopt a two-level model to describe the results for the below-bandgap excitations.
The relaxation dynamics of a two-level system is described by the following set of equations
\begin{align}
\label{func1}
\frac{\text{d}C(t)}{\text{d}t} &= +p(t)-\frac{C(t)}{\tau_1} \\
\label{func2}
\frac{\text{d}G(t)}{\text{d}t} &= -p(t)+\frac{C(t)}{\tau_1},
\end{align}
with $G(t)$, $C(t)$, and $p(t)$ being the population of the ground state, the excited state, and a Gaussian pulse, respectively, and $\tau_1$ the time constant.
The change $\Delta E$ is assumed to be proportional to the population of the excited state\cite{Ulbricht2011}. 
As one can see, the two-level-system description fits the relaxation of the signal well and the rise time of the signal is well described with an excitation with a Gaussian pulse (see dashed lines in Figs.~\ref{fig2} and  \ref{fig4}).
In Fig.~\ref{fig6} (b) we show the relaxation time extracted from the \textit{n}- and \textit{p}-type samples. 
The \textit{p}-type sample with a relaxation time of \SI{3}{ps} relaxes twice as fast as the \textit{n}-type sample with a relaxation time of \SI{6}{ps}, at the highest fluence of \SI{27}{\micro J/cm^2}.
These relaxation times, in the order of a few ps, are often found for the more metallic TIs in literature\cite{Choi2018,Cheng2014}, and this timescale is typically attributed to phonon mediated relaxation.
The difference in amplitude and relaxation time for the two samples could be due to the following difference of the underlying physical properties.
One possibility is that the effective mass of the charge carrier reported for the \textit{n}- and \textit{p}- type TIs are different\cite{Witting2020, Taskin2011}, resulting in different scattering rates.
Another one is that the underlying excitation is a process involving states inside the bandgap, like the surface state, impurity bands or charge carrier puddles\cite{Borgwardt2016,Tang2013,Zhang2011}.\\
\indent The above-bandgap excitation measurements show a clear deviation from an exponential decay (see Fig. \ref{fig3} and \ref{fig5}), therefore the corresponding relaxation dynamics needs a different description than the exponential decay.
The compressed exponential function was chosen because it fits well to the deviation from the exponential decay, that we observed in all the above-bandgap measurements (see dashed lines in Fig. \ref{fig3} and \ref{fig5}).
The compressed exponential function is defined as
\begin{align}
\label{func3}
\Delta E/E = Ae^{-\left(\frac{\tau-\tau_0}{\tau_1}\right)^\alpha},
\end{align}
with $A$, $\tau_0$, $\tau_1$, and $\alpha$ being the amplitude, peak position, relaxation time, and compression parameter, respectively.
The amplitude $A$ is set to be the maximum change of $-\Delta E/E$, with $\tau_0$ being the pump-probe delay corresponding to the maximum.
The fitting results are summarized in Fig.~\ref{fig7} as a function of pump fluence.
As shown in Fig.~\ref{fig7} (a), the maximum change of $-\Delta E/E$ in the \textit{p}-type sample exhibits a linear dependence on the fluence at least up to \SI{4}{\micro J/cm^2}, while for the \textit{n}-type sample a saturation-like behavior is observed toward higher fluence above \SI{4}{\micro J/cm^2}.\\
\indent In Fig.~\ref{fig7} (b) one can see that the relaxation time increases from \SI{1.5}{ps} at fluences of $\approx \SI{0.6}{\micro J/cm^2}$ to \SI{6.2}{ps} at fluences of \SI{5.8}{\micro J/cm^2} for the \textit{p}-type BST.
The relaxation time of the \textit{n}-type BST increases as well from \SI{2.5}{ps} at fluences of $\approx \SI{0.6}{\micro J/cm^2}$ to \SI{7}{ps} at fluences of \SI{7.8}{\micro J/cm^2}.
The increase in relaxation time with increasing fluence is a dependence often observed in metals, which hints to a metallic nature of the electron dynamics\citep{Choi2018}.
This metallic nature of the electron dynamics has been shown to exist in other TIs\cite{Cheng2014}.
In combination with the \si{ps} timescale of the relaxation time this indicates that the electron relaxation is predominantly through the scattering with phonons \cite{Groeneveld1995}.
Furthermore, the compression parameter with $\alpha = 1.6$ is nearly constant over the whole measured range fluence range from \SI{0.6}{} to \SI{7.7}{\micro J/cm^2}.
Typically, compressed exponential functions are used to describe a system that deviates from the normal exponential decay\cite{Whitehead2009}.
This occurs when the underlying processes get more complex than a simple relaxation from one independent state to another\cite{Whitehead2009} (see Appendix for an illustration). 
In the case of BST this can be due to the interaction of electrons in the conduction band with states inside the bandgap.
Probable candidates for these are the surface states,  impurities, or charge carrier puddles\cite{Choi2018,Borgwardt2016}.
The fact that the compression parameter is independent of the fluence shows that the interaction between the excited states is independent of the number of excited electrons.\\
\indent For the temperature dependence measurement we use the compressed exponential function as well, because at a fluence of \SI{1}{\micro J/cm^2} we still observed an deviation from normal exponential decay (see Fig.~\ref{fig5}).
Figure \ref{fig8} shows all extracted fitting parameters from the \textit{n}-type and \textit{p}-type samples.
In Fig. \ref{fig8}(a), Max($-\Delta E/E$) for both samples decreases monotonically with increasing temperature over the whole range from \SI{5}{K} to \SI{300}{K}. 
The difference between the samples is, that the \textit{p}-type sample has the smaller signal and slope.
We speculate that the difference in signal is due to higher number of charged carriers of the \textit{p}-type sample in its equilibrium state shown by the Hall measurements in Fig. \ref{fig1} (b).
With the charge carrier densities extracted from the Hall measurements and effective masses in Refs.~\cite{Witting2020, Taskin2011}, we estimate a plasma frequency of $\approx \SI{50}{THz}$(\SI{70}{THz}) and $\approx \SI{134}{THz}$(\SI{72}{THz}) at \SI{2}{K}(\SI{300}{K}) for the \textit{n}- and \textit{p}-type BST, respectively.
The difference in signal at low temperatures therefore can be explained by a larger relative increase in plasma frequency upon pump excitation.
Another indicator for the influence of the charge carrier density is the smaller difference at higher temperature, here the plasma frequency is almost identical.
Figure~\ref{fig8}(b) shows that the relaxation time increases slightly with increasing temperature, which is similar to the behavior of metals in pump-probe experiments.
This dependence can be described with the two-temperature-model\cite{Groeneveld1995}, in the same way as the fluence dependence of the relaxation time.
This further hints at the metallic nature of the relaxation in the above-bandgap excitation.
A similar behavior was found in measurements in the TI compound Bi$_{1.5}$Sb$_{0.5}$Te$_{1.8}$Se$_{1.2}$\cite{Cheng2014}.

\section{Summary}

To summarize, we performed time-resolved mid-infrared-pump terahertz-transmission-probe study of nonequilibrium charge-carrier dynamics in the \textit{n}-type and the \textit{p}-type topological insulators (Bi$_{1-\text{x}}$Sb$_{\text{x}}$)$_2$Te$_3$.
The relaxation dynamics of charge carriers was investigated as a function of temperature and pump fluence, for pump-pulse energies below and above the bandgap in the two types of samples.
While for the below-bandgap excitation the relaxation dynamics exhibits an exponential decay signalling a noninteracting process, the above-bandgap excitation leads to a compressed exponential decay reflecting a more complex relaxation process involving interactions of different states in the out-of-equilibrium system.
These states are very likely the in-gap states, e.g. surface states, impurity states and charge puddles, as  
a saturation behaviour in the fluence dependence measurement was observed for both the \textit{n}-type and the \textit{p}-type samples.
In addition, the observed difference in temperature dependence for the two types of samples points to the  effects of equilibrium-state charge carrier density on the relaxation dynamics.

\begin{acknowledgments}
We acknowledge partial supported by
the DFG German Research Foundation) via Project No. 277146847 — Collaborative Research Center 1238: Control and Dynamics of Quantum Materials (Subprojects No. A04, B05). G.L. acknowledges the support by the Research Foundation — Flanders (FWO, Belgium), project Nr. 27531 and 52751.
\end{acknowledgments}

\appendix*
\section{}
\begin{figure}[h]
        \center{\includegraphics[width=0.45\textwidth]
        {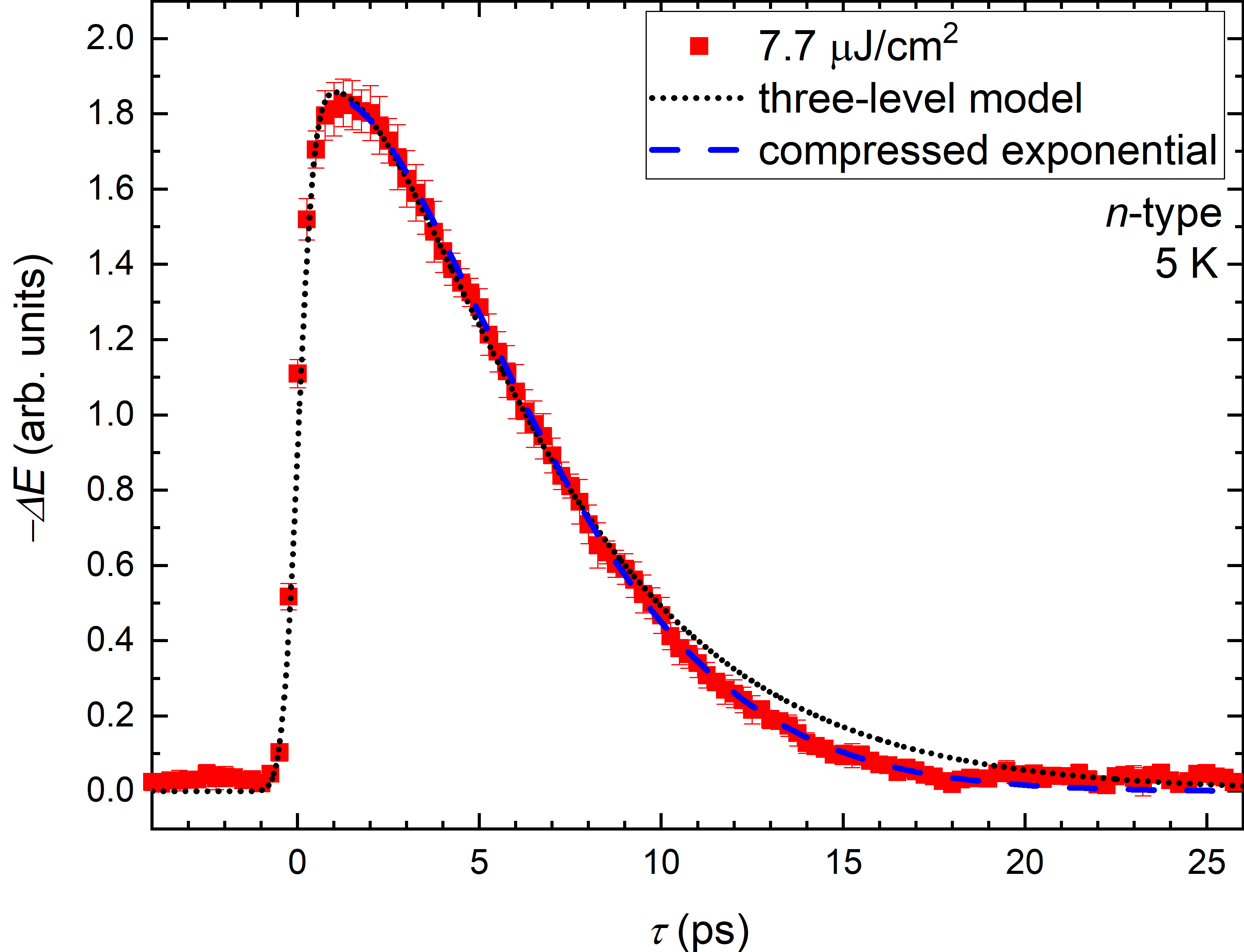}}
        \caption{\label{fig9} Red squares: \textit{Above-bandgap} excitation measurement of the \textit{n}-type sample at \SI{5}{K} and \SI{7.7}{\micro J/cm^2}. Dashed line: Compressed exponential fit corresponding to the data (see Fig.~\ref{fig3}). Dotted line: three-level model simulation.}
\end{figure}
To illustrate that the deviation of the relaxation dynamics from an exponential decay can be a result of interactions with an intermediate state, we present a simulation based on a three-level model system. In comparison to the two-level system in Eqs.~(\ref{func1}) and (\ref{func2}), we consider relaxation processes through an additional state $S(t)$ as follows:
\begin{align}
\frac{\text{d}C(t)}{\text{d}t} &= +p(t)-\frac{C(t)}{\tau_{CS}}-\frac{C(t)}{\tau_{CG}} \\
\frac{\text{d}S(t)}{\text{d}t} &= +\frac{C(t)}{\tau_{CS}}-\frac{S(t)}{\tau_{SG}}\\
\frac{\text{d}G(t)}{\text{d}t} &= -p(t)+\frac{C(t)}{\tau_{CG}}+\frac{S(t)}{\tau_{SG}}.
\end{align}
As in Eqs.~(\ref{func1}) and (\ref{func2}), $G(t)$ and $C(t)$ describe the population of the ground state and conduction band, respectively, while $p(t)$ describes the excitation function.
Additionally, $S(t)$ describes the population of the intermediate state with $\tau_{CS}$, $\tau_{CG}$, and $\tau_{SG}$ being the time constants for the transitions between conduction band and intermediate state, conduction band and ground state, and intermediate state and ground state, respectively. Assuming the change of the electric field only depends on the population of the conduction band and the intermediate state, we adopt a constant $\beta$ to weight the different contributions\cite{Choi2018}:
\begin{align}
\Delta E(t) = C(t) + \beta S(t).
\end{align}
As shown in Fig.~\ref{fig9}, this three-level model can be applied to describe the relaxation dynamics of the \textit{n}-type sample at \SI{5}{K} and \SI{7.7}{\micro J/cm^2}, where we use an Gaussian-type excitation pulse with a temporal width (FWHM) of \SI{800}{fs}, the time constants $\tau_{CS} = \SI{2.8}{ps}$, $\tau_{SG} = \SI{4.4}{ps}$, $\tau_{CG} = \SI{30}{ps}$, and the weight $\beta = 1.4$. Although the description of the experimental data based on the phenomenological three-level model is not as perfect as the compressed exponential decay function, this simulation clearly indicates the involvement of more complex relaxation processes.

\bibliographystyle{apsrev4-1}
\bibliography{BST_bib}

\end{document}